\documentstyle[preprint,aps]{revtex}

\tighten
\begin{document}

\title{
       Spurious states in the Faddeev formalism for few-body systems
}
\medskip

\author{
        P. Navr\'atil\footnote{On leave of absence from the
   Institute of Nuclear Physics,
                   Academy of Sciences of the Czech Republic,
                   250 68 \v{R}e\v{z} near Prague,
                     Czech Republic.}$^a$,
     B. R. Barrett$^a$, and W. Gl\"ockle$^b$
        }

\medskip

\address{
                   $^a$Department of Physics,
                   University of Arizona,
                   Tucson, Arizona 85721\newline
                   $^b$Ruhr-Universit\"at Bochum, D-44780 Bochum,
                       Germany
}

\maketitle

\bigskip

\begin{abstract}
We discuss the appearance of spurious solutions of few-body
equations for Faddeev amplitudes. The identification of spurious
states, i.e., states that lack the symmetry required for solutions
of the Schr\"odinger equation, as well as the symmetrization of the
Faddeev equations is investigated. As an example, systems of three and 
four electrons, bound in a harmonic-oscillator potential and 
interacting by the Coulomb potential, are presented.
\end{abstract}

\bigskip
\bigskip

\narrowtext



\section{Introduction}
\label{sec1}

One of the most viable approaches for solving the 
few-body problem is the Faddeev method \cite{Fad60}.
It has been successfully applied to solve the three-nucleon bound-state
problem for various nucleon-nucleon 
potentials \cite{PFGA80,PFG80,CPFG85,FPSS93}.
The most complex calculations of this kind include
up to 82 channels, when all the $j\leq 6$ waves are taken
into account\cite{NHKG97}.

The Hamiltonian in the Schr\"odinger equation
is Hermitian and the solutions for a system of fermions, for example,
are antisymmmetrized. On the other hand, 
the Faddeev equations are non-Hermitian and the trial
wave functions used for a system of 
fermions, e.g., are not fully antisymmetrized. 
Also, there are three Faddeev equations for the three-body system
acting on the same variables as in the Schr\"odinger equation.
One, therefore, could expect that there should be three times
as many solutions as for the Schr\"odinger equation. 
Then, it is not surprising that spurious solutions 
of the Faddeev equations exist.  
A spurious solution is an eigenstate,
which does not have the symmetry required by the Schr\"odinger
equation, e.g., antisymmetry for a system of identical fermions. 
A  spurious component in a solution of 
the Faddeev equations was found analytically for the first time 
for the ground-state of 
three identical particles bound in the harmonic-oscillator (HO) 
potential by J. L. Friar {\it et al.} \cite{FGP80}.
Spurious components of the Faddeev amplitudes
were then observed also for the excited states
in extensions of this work for three identical particles \cite{BM92}, 
three nonidentical particles \cite{BM94}, as well as for four
identical spinless particles \cite{ZM94}. 
Spurious solutions of Faddeev-like equations were investigated
by Evans and Hoffman \cite{EH81}, and the
existence of spurious solutions of Faddeev equations for three
identical particles
was recently demonstrated by Rudnev and Yakovlev \cite{RY95}. In Ref.
\cite{RY95} such solutions were constructed for states of zero total
angular momentum. 
In addition, spurious solutions of the Faddeev equations 
for three nonidentical
particles interacting by central potentials were investigated
by V. V. Pupyshev \cite{P96}.
In Ref. \cite{NB98}, we noted the appearance of spurious solutions,
whose number exceeded the number of physical solutions,
of the Faddeev equations for a three-nucleon system solved
in an HO basis in a shell-model approach.
Also, a spurious solution was reported recently in 
a three-body model calculation of $^9_\Lambda$Be \cite{OKSYN98}.  

In the present paper we investigate the appearance of spurious
states in the Faddeev formalism and their identification
in a systematic manner. 
We use three- and four-electron systems bound in an HO
potential as an example for illustration and quantification of the problem.

In section \ref{sec2} we discuss the three fermion system.  
A generalization for the four-body system is presented in
section \ref{sec3}. Conclusions are given in section \ref{sec4}.

\section{Three-body system}
\label{sec2}

Our discussion is quite general. 
However, we prefer to illustrate our points by using a particular simple
example, namely, a system of electrons bound by an 
HO potential and interacting by the Coulomb potential.
We consider, therefore, the following Hamiltonian
\begin{equation}\label{hamomega}
H=\sum_{i=1}^A \left[ \frac{\vec{p}_i^2}{2m}
+\frac{1}{2}m\Omega^2 \vec{r}^2_i
\right] + \sum_{i<j}^A V_{\rm C}(\vec{r}_i-\vec{r}_j) \; .
\end{equation}
Eigenstates of the Hamiltonian (\ref{hamomega}) are antisymmetric with 
respect to the exchange of any electron pair.

In the Faddeev formalism for a system of three identical particles, 
i.e., $A=3$, the following transformation of the coordinates 
\begin{mathletters}\label{rtrans}
\begin{eqnarray}
\vec{r}&=&\sqrt{\textstyle{\frac{1}{2}}}(\vec{r}_1-\vec{r}_2) \; ,
\\
\vec{y}&=&\sqrt{\textstyle{\frac{2}{3}}}[\textstyle{\frac{1}{2}}(\vec{r}_1
+\vec{r}_2)-\vec{r}_3] \; ,
\end{eqnarray}
\end{mathletters}
and, similarly, of the momenta, is introduced. It
brings the one-body HO Hamiltonian from Eq. (\ref{hamomega})
into the form
\begin{equation}\label{H0}
H_0 =  \frac{\vec{p}^2}{2m} + \frac{1}{2}m\Omega^2 \vec{r}^2
     + \frac{\vec{q}^2}{2m} + \frac{1}{2}m\Omega^2 \vec{y}^2 \; ,
\end{equation}
with the trivial center-of-mass term omitted.
Eigenstates of $H_0$, 
\begin{equation}\label{hobas}
|n l s j, {\cal N} {\cal L} {\cal J},  J \rangle \; ,
\end{equation}
can be used as the basis for the Faddeev calculation.
Here $n, l$ and ${\cal N}, {\cal L}$ are the HO
quantum numbers
corresponding to the harmonic oscillators associated with 
the coordinates and 
momenta $\vec{r}, \vec{p}$ and $\vec{y}, \vec{q}$, respectively. 
The quantum numbers $s,j$ describe the spin and angular momentum
of the relative-coordinate partial channel of particles 1 and 2, 
${\cal J}$ is the angular momentum of the third particle 
relative to the
center of mass of particles 1 and 2 and $J$ is the total angular 
momentum. It is obvious that the basis (\ref{hobas}) 
(similarly for trial states used in traditional Faddeev calculations), 
while antisymmetrized
with respect the exchange of particles $1\leftrightarrow 2$,
is not antisymmetrized
with respect the exchanges of particles $1\leftrightarrow 3$ 
and $2\leftrightarrow 3$. On the other
hand, the physical solutions corresponding to the solutions of the 
Schr\"odinger equation must be totally antisymmetrized. 
For our example system, the states (\ref{hobas}) form a complete orthonormal
basis. The physical solutions are such linear combinations of the basis
states that have the proper antisymmetry for all the particle exchanges.
On the other hand, one must expect that there are more basis states
(\ref{hobas}) than the total possible number of linearly independent
antisymmetrized linear combinations of the states (\ref{hobas}).  
Consequently, it is 
natural that spurious states will
appear, when the calculation is performed using a basis of the type
(\ref{hobas}).

The first Faddeev equation can be written in the differential form as
\begin{equation}\label{faddif1}
( H_0 + V_3  - E ) \phi_3  = - V_3 (\phi_1 
+ \phi_2 ) \; ,
\end{equation}
where $V_3$ is the two-body interaction between particles  
1 and 2 and $\phi_1, \phi_2$, and $\phi_3$ are
the three Faddeev components. The other two Faddeev equations 
are cyclical versions of the Eq. (\ref{faddif1}).
Equation (\ref{faddif1}) can be re-written in the form
\begin{equation}\label{Fadeq}
\tilde{H}|\psi\rangle_{\rm K} = E|\psi\rangle_{\rm K} \; ,
\end{equation}
with
\begin{equation}\label{Fadham}
\tilde{H}= H_0 + V(\vec{r}) {\cal T} \; .
\end{equation}
Here, $V(\vec{r})$ is the interaction between particles 1 and 2,
e.g., $V_3$ of Eq. (\ref{faddif1}) or $V_{\rm C}(\sqrt{2}\vec{r})$ 
of Eq. (\ref{hamomega}); $|\psi\rangle_{\rm K}$ corresponds to the 
Faddeev amplitude $\phi_3$ in Eq. (\ref{faddif1});
and ${\cal T}$, which has the properties of 
a metric operator \cite{PFG80,SGH}, is given by
\begin{equation}\label{metric}
{\cal T}=1+{\cal T}^{(-)}+{\cal T}^{(+)} \; ,
\end{equation}
with ${\cal T}^{(+)}$ and ${\cal T}^{(-)}$ the cyclic and the anticyclic 
permutation operators, respectively. 

For the basis (\ref{hobas}), which we use as an example 
in the present paper, we obtain a formula for the matrix elements of 
${\cal T}^{(-)}+{\cal T}^{(+)}$ by simplification of the expression
(10) in Ref. \cite{NB98}, namely
\begin{eqnarray}\label{t13t23}
&&\langle n_1 l_1 s_1 j_1, {\cal N}_1 {\cal L}_1  
{\cal J}_1, J  | {\cal T}^{(-)}+{\cal T}^{(+)} |  
n_2 l_2 s_2 j_2, {\cal N}_2 {\cal L}_2  
{\cal J}_2, J\rangle = \delta_{N_1,N_2} \nonumber \\
&& \times \sum_{LS} \hat{L}^2 \hat{S}^2
\hat{j}_1 \hat{j}_2 \hat{\cal J}_1 \hat{\cal J}_2 \hat{s}_1 \hat{s}_2
(-1)^L
          \left\{ \begin{array}{ccc} l_1   & s_1   & j_1   \\
          {\cal L}_1  & \textstyle{\frac{1}{2}}  & {\cal J}_1 \\
                                     L   & S  & J
\end{array}\right\}
          \left\{ \begin{array}{ccc} l_2   & s_2   & j_2   \\
       {\cal L}_2  & \textstyle{\frac{1}{2}}   & {\cal J}_2 \\
                                     L   & S  & J
\end{array}\right\}
  \left\{ \begin{array}{ccc} \textstyle{\frac{1}{2}} & \textstyle{\frac{1}{2}} 
               & s_1 \\
             \textstyle{\frac{1}{2}}  &  S  & s_2
\end{array}\right\}
\nonumber \\
&&\times \left[(-1)^{s_1+s_2-{\cal L}_1-l_1} 
\langle {\cal N}_1 {\cal L}_1 n_1 l_1 L 
| n_2 l_2 {\cal N}_2 {\cal L}_2 L \rangle_{\rm 3}
+\langle n_1 l_1 {\cal N}_1 {\cal L}_1 L 
| {\cal N}_2 {\cal L}_2 n_2 l_2 L \rangle_{\rm 3}
\right] \; ,
\end{eqnarray}
where $N_i=2n_i+l_i+2{\cal N}_i+{\cal L}_i, i\equiv 1,2$, 
$\hat{j}=\sqrt{2j+1}$ 
and $\langle {\cal N}_1 {\cal L}_1 n_1 l_1 L 
| n_2 l_2 {\cal N}_2 {\cal L}_2 L \rangle_{\rm 3}$
is the general HO bracket for two particles with mass 
ratio 3 as defined, e.g., in Ref. \cite{Tr72}.
Similar expressions for different bases are described, e.g., 
in Refs. \cite{HKT72,Glo83}. It follows from the symmetry properties
of the states (\ref{hobas}) and of the HO brackets
that the contributions of ${\cal T}^{(-)}$ and ${\cal T}^{(+)}$
in (\ref{t13t23}) are identical.

We note that the eigensystem of the operator ${\cal T}$ (\ref{metric}) 
consists
of two subspaces. The first subspace has eigenstates with the eigenvalue
3, which form totally antisymmetric physical states, while the second 
subspace has eigenstates with
the eigenvalue 0, which form a not completely antisymmetric, unphysical 
subspace of states. Although we found these properties of ${\cal T}$
by direct calculation using the relation (\ref{t13t23}), it is, in fact,
a general result. The same structure of eigenstates was also reported
in Ref. \cite{RY95} using a different basis. The eigenvalue structure
follows from the fact that $\frac{1}{3}{\cal T}$ has the properies of
a projection operator.

The Hamiltonian $\tilde{H}$ (\ref{Fadham}) is non-Hermitian. 
By solving the equation (\ref{Fadeq}) one obtains the
right (ket) eigenstates, while by acting with $\tilde{H}$
to the left, one gets the bi-orthogonal left (bra) eigenstates.     
In the basis
with physical and spurious states separated, e.g., those obtained
from the diagonalization of ${\cal T}$, the Hamiltonian
matrix takes the form
\begin{equation}\label{hamstruct}
\left( 
\begin{array}{cc} 
\langle {\rm Ph}|\tilde{H}|{\rm Ph}\rangle & 0 \\
\langle {\rm Sp}|\tilde{H}|{\rm Ph}\rangle & 
\langle {\rm Sp}|\tilde{H}|{\rm Sp}\rangle 
\end{array}
\right)  \; .
\end{equation}
Formally we obtain the Hamiltonian matrix (\ref{hamstruct}) from the
Hamiltonian matrix in the basis of the type (\ref{hobas}) by an orthogonal
transformation that transforms the basis (\ref{hobas}) into the
eigenstates of the operator ${\cal T}$ (\ref{metric}).
In fact, this situation is very much analogous to the
well-known properties of the Dyson boson-mapped systems \cite{Dy56}.
It follows from the structure of the Hamiltonian matrix 
(\ref{hamstruct}) that its right spurious eigenstates 
$|\psi; {\rm Sp}\rangle_{\rm K}$ 
are not contaminated by the physical states. Similarly, the
left physical eigenstates $_{\rm B}\langle \psi; {\rm Ph}|$ 
are not contaminated by the spurious ones. On the other hand,
its right physical eigenstates $|\psi; {\rm Ph}\rangle_{\rm K}$
may have spurious admixtures; similarly the left spurious
eigenstates $_{\rm B}\langle \psi; {\rm Sp}|$ may have physical
admixtures. 
As a right spurious eigenstate has no physical
admixtures, it must be annihilated by the action of the
operator ${\cal T}$ (\ref{metric}), e.g.,
\begin{equation}\label{anihil}
{\cal T}|\psi; {\rm Sp}\rangle_{\rm K} = 0 \; .
\end{equation}
The action of the opeartor ${\cal T}$ on the right eigenstates,
thus, serves as a test
for identification of spurious states among the calculated
eigenstates, in analogy to identification of spurious
states in the Dyson boson mapping by the means of the (so-called)
${\cal R}$ operator \cite{Rproj}. The relation (\ref{anihil})
can be expressed in terms of the Faddeev amplitudes appearing
in Eq. (\ref{faddif1}) as $\phi_1+\phi_2+\phi_3=0$.
We note that the sum of the three Faddeev components has to be indentified
with the solution of the Schr\"odinger equation.
If one finds solutions of the Faddeev equations, which are 
unphysical, e.g., that do not have the proper symmetry with respect
to the particle exchanges, the corresponding three Faddeev components have 
to add up to zero. This is the only possibilty to avoid a contradiction. 

It is possible to avoid the spurious state problem completely by
hermitizing the 
Hamiltonian (\ref{Fadham}) on the physical subspace, where it is 
quasi-Hermitian (see the discussion of quasi-Hermitian 
operators, e.g., in Ref. \cite{SGH}).
The Hermitized Hamiltonian takes the form
\begin{equation}\label{Fadhamh}
\bar{H}= H_0 + \bar{{\cal T}}^{1/2}V(\vec{r})\bar{{\cal T}}^{1/2} \; ,
\end{equation}
where $\bar{{\cal T}}$ operates on the physical subspace only,
e.g., it has only the eigenvalues 3.

Let us mention several relations between the eigenstates of the
Faddeev Hamiltonian $\tilde{H}$ (\ref{Fadham}) and the
symmetrized Hamiltonian $\bar{H}$ (\ref{Fadhamh}). First,
we have for the left and right eigenstates of $\tilde{H}$  
\begin{equation}\label{KBrel}
_{\rm B}\langle \psi; {\rm Ph}| = \;\;_{\rm K}\langle \psi; {\rm Ph}|
{\cal T} \; .
\end{equation}
The eigenvectors $|\Psi\rangle$ of the Hamiltonian 
$\bar{H}$ (\ref{Fadhamh}) are related
to the eigenvectors of the Hamiltonian $\tilde{H}$ (\ref{Fadham}) by
\begin{mathletters}
\begin{eqnarray}
|\Psi\rangle &=& {\cal T}^{1/2} |\psi; {\rm Ph}\rangle _{\rm K} \; ,  \\
\langle\Psi| &=& _{\rm B}\langle\psi; {\rm Ph}|\bar{{\cal T}}^{-1/2} \; .
\end{eqnarray}
\end{mathletters}
Note that we write explicitly $\bar{{\cal T}}$ instead of ${\cal T}$
only when an inversion is needed. 
For matrix elements of a general operator $O$ we then have
\begin{equation}\label{O1}
\langle \Psi_f | O | \Psi_i \rangle = \;\;
_{\rm B}\langle \psi_f; {\rm Ph}|\bar{{\cal T}}^{-1/2} O
{\cal T}^{1/2} |\psi_i; {\rm Sp}\rangle _{\rm K} \; . 
\end{equation}

In addition to the relation (\ref{anihil}), there is
another possibility of identification of spurious states.
It follows from the properties of the matrix elements
of a general operator $O$ between the physical and spurious
eigenstates of $\tilde{H}$ (\ref{Fadham}), namely
\begin{mathletters}\label{O2}
\begin{eqnarray}
_{\rm B}\langle \psi_f; {\rm Ph}|\bar{{\cal T}}^{-1/2} O
{\cal T}^{1/2} |\psi_i; {\rm Sp}\rangle_{\rm K} &=& 0 \; ,  \\
_{\rm B}\langle \psi_f; {\rm Sp}|\bar{{\cal T}}^{-1/2} O
{\cal T}^{1/2} |\psi_i; {\rm Ph}\rangle_{\rm K} &\neq & 0 \; .  
\end{eqnarray}
\end{mathletters}
These relations follow from the properties of the left and right 
eigenstates as discussed in the paragraph after Eq. (\ref{hamstruct}).
For an operator that commutes with ${\cal T}$ the square
roots of ${\cal T}$ may be omitted in Eqs. (\ref{O1},\ref{O2}).
Also, in the above relations the substitution 
$\sum_{i<j}O_{ij}={\cal T}^{1/2} O(\vec{r}) {\cal T}^{1/2}$ can be used 
for a two-body operator $O_{ij}$ depending on $(\vec{r}_i-\vec{r}_j)$.

Let us return to our specific system example.
The metric ${\cal T}$ (\ref{metric}) is diagonal in 
$N=2n+l+2{\cal N}+{\cal L}$. It follows from the expression (\ref{t13t23})
that any basis truncation other than one of the type
$N\le N_{\rm max}$ would lead, in general, to mixing of 
physical and unphysical states. 
Here, $N_{\rm max}$ characterizes the maximum of total allowed 
harmonic-oscillator quanta in the basis.
At the same time, the truncation into total allowed
oscillator quanta $N\le N_{\rm max}$ preserves the equivalence of the
Hamiltonians (\ref{Fadham}) and (\ref{Fadhamh}) on the physical subspace.
A general consequence of this observation is that an improper
truncation in the treatment of ${\cal T}$ leads to physical
and spurious state mixing.  

In Table \ref{tab1} we present the dimensions (D) of the basis (\ref{hobas})
corresponding to a particular $N=2n+l+2{\cal N}+{\cal L}$
together with the number of physical states (Ph) and spurious states
(Sp) for the three-electron system with $J^\pi=1/2^-$. 
Apparently, the number of physical states is about  a third of all
basis states.
We also present
the ground state and the first excited state energies obtained 
with the basis restricted by $N\le N_{\rm max}$,
where $N_{\rm max}$ corresponds to the number in the first row.
We used the HO energy $\hbar\Omega=0.5$ 
atomic units (a.u.).
The physical eigenenergies are shown without the trivial
center-of-mass contribution. 
It is immediately seen from 
Eqs. (\ref{Fadham}) and (\ref{anihil}) that the spurious states
have the energies corresponding to the unperturbed Hamiltonian 
$H_0$ (\ref{H0}).
Typically in the search for a bound state 
in the Faddeev calculations the lowest state would be physical 
and the spurious state existence would be unnoticed. 
The present electron
system with a repulsive Coulomb interaction is interesting 
because of the fact that the physical ground state
is the tenth state as can be deduced from the Table \ref{tab1}.
There are 2 spurious states with the unperturbed HO energy
2 a.u. and 7 spurious states with the energy 3 a.u. 
We note that for the discussed system the ground-state energy
obtained by the Stochastic Variational Method (SVM) \cite{Varga1}
is 3.26324 a.u. \cite{Varga2} (after subtracting 0.75 a.u. for 
the center-of-mass energy). In our HO basis
calculation two-decimal place precision is obtained rather
rapidly. A further improvement of the precision is, 
however, slow. We performed calculations up to $N_{\rm max}=39$,
where we obtained the ground state energy of 3.2634 a.u. and
4.0446 a.u. for the first-excited state. A substantial
acceleration of convergency can be achieved by employing the
effective interaction approach in a manner similar to that 
discussed in Ref. \cite{NB98}. By replacing the interaction 
$V(\vec{r})$ in Eqs. (\ref{Fadham},\ref{Fadhamh}) 
by $V_{\rm eff}(\vec{r})$ we reach the SVM ground-state
result for $N_{\rm max}=27$ and the for the first excited state 
we then obtain 4.04458 a.u.

\section{Four-body system}
\label{sec3}

To demonstrate that the problems discussed 
prevail also for the Faddeev-type approach to systems with more than
three particles, we present briefly the extension of the studied 
system to four electrons. We use the Hamiltonian 
(\ref{hamomega}) with $A=4$. By introducing 
the coordinate (and momenta) transformation
\begin{mathletters}\label{fourtran}
\begin{eqnarray}
\vec{r}&=&\sqrt{\textstyle{\frac{1}{2}}}(\vec{r}_1-\vec{r}_2) \;, 
\\
\vec{y}&=&\sqrt{\textstyle{\frac{2}{3}}}[\textstyle{\frac{1}{2}}(\vec{r}_1
+\vec{r}_2)-\vec{r}_3] \;, 
\\
\vec{z}&=&\textstyle{\frac{\sqrt{3}}{2}}[\textstyle{\frac{1}{3}}(\vec{r}_1
+\vec{r}_2+\vec{r}_3)-\vec{r}_4] \;, 
\end{eqnarray}
\end{mathletters}
the one-body part 
of the Hamiltonian (\ref{hamomega}) is obtained as
\begin{equation}\label{H0four}
H_0 =  \frac{\vec{p}^2}{2m} + \frac{1}{2}m\Omega^2 \vec{r}^2
     + \frac{\vec{q}^2}{2m} + \frac{1}{2}m\Omega^2 \vec{y}^2 
     + \frac{\vec{o}^2}{2m} + \frac{1}{2}m\Omega^2 \vec{z}^2 \; ,
\end{equation}
with the trivial center-of-mass term omitted.

A possible generalization of the Faddeev equation (\ref{Fadeq})
for four identical particles can be written in the form  
\begin{equation}\label{fadfour}
\tilde{H}|\psi_{(123)4}\rangle = E|\psi_{(123)4}\rangle \; ,
\end{equation}
with
\begin{equation}
\tilde{H}|\psi_{(123)4}\rangle \equiv  H_0 |\psi_{(123)4}\rangle 
+ \textstyle{\frac{1}{2}}(V_{12}+V_{13}+V_{23})
(|\psi_{(123)4}\rangle 
+|\psi_{(432)1}\rangle +|\psi_{(134)2}\rangle 
+|\psi_{(142)3}\rangle )\; ,
\end{equation}
and
\begin{equation}\label{metric4}
(|\psi_{(123)4}\rangle 
+|\psi_{(432)1}\rangle +|\psi_{(134)2}\rangle 
+|\psi_{(142)3}\rangle ) =
(1-{\cal T}_{14}-{\cal T}_{24}-{\cal T}_{34})|\psi_{(123)4}\rangle 
\equiv {\cal T}_4 |\psi_{(123)4}\rangle \; .
\end{equation}
Here, $|\psi_{(123)4}\rangle $ is a four-fermion Faddeev amplitude
completely antisymmetrized for particles 1,2, and 3. There are 
three other equations that can be obtained from Eq. (\ref{fadfour}) by
permuting particle 4 with particles 1, 2, and 3. Their sum then leads
to the Schr\"odinger equation. We note that the present equations
are different from the traditional Faddeev-Yakubovsky equations
\cite{Ya67}, which combine Faddeev amplitudes depending on two
sets of Jacobi coordinates. As we are working with a complete orthonormal
basis, it is sufficient and convenient to use a single set
of Jacobi coordinates defined by the relations (\ref{fourtran}).
Unlike the Faddeev amplitudes used typically in the Faddeev-Yakubovsky
equations, the amplitudes appearing in Eq. (\ref{fadfour}) are 
antisymmetrized with respect to the first three particles.
The present equations allow to employ easily three-body interactions
or three-body effective interactions. The latter property makes 
them particularly
useful for an extension of shell-model calculations for four nucleons
\cite{NB99}.

The starting point for the present four-electron calculation
is the basis
\begin{equation}\label{fourbas}
|N_1 i J_1, n_z l_z {\cal J}_4, J \rangle \; ,
\end{equation}
with the three-fermion part given by
the antisymmetrized eigenstates of ${\cal T}$ (\ref{metric}) 
corresponding to eigenvalue 3, e.g., 
\begin{equation}
|N_1 i J_1 \rangle =
\sum c_{n l s j {\cal N} {\cal L}
{\cal J}_{3}}^{N_{1 } i J_{ 1} }
|n l s j , {\cal N} {\cal L} {\cal J}_3, J_1 \rangle
\; ,
\end{equation}
where $N_1=2n+l+2{\cal N}+{\cal L}$ and $i$ counts the eigenstates
of ${\cal T}$ with the eigenvalue 3 for given $N_1$ and $J_1$.

As for the three-particle transposition operators
(\ref{t13t23}),
a compact formula can be derived for the matrix elements
of the four-particle transposition operators in the basis
(\ref{fourbas}), e.g.,
\begin{eqnarray}\label{t4}
&&\langle N_{\rm 1L} i_{\rm L} J_{\rm 1L} , n_{z \rm L}
l_{z \rm L} {\cal J}_{\rm 4L}, J | {\cal T}_{14}
+{\cal T}_{24}+{\cal T}_{34}
|N_{\rm 1R} i_{\rm R} J_{\rm 1R}, n_{z \rm R}
l_{z \rm R} {\cal J}_{\rm 4R}, J \rangle \nonumber \\
&&= - \delta_{N_{\rm L},N_{\rm R}} \sum
c_{n_{\rm L} l_{\rm L} s_{\rm L} j_{\rm L}  
{\cal N}_{\rm L} {\cal L}_{\rm L}
{\cal J}_{3\rm L}}^{N_{1 \rm L} i_{\rm L} J_{\rm 1L}}
c_{n_{\rm R} l_{\rm R} s_{\rm R} j_{\rm R}  
{\cal N}_{\rm R} {\cal L}_{\rm R}
{\cal J}_{3\rm R}}^{N_{1 \rm R} i_{\rm R} J_{\rm 1R}}
\hat{L}_{\rm 1L}^2 \hat{L}_{\rm 1R}^2 \hat{S}_{\rm 1L}^2
\hat{S}_{\rm 1R}^2 \hat{L}_2^2 \hat{S}_2^2
\nonumber \\
&&\times \hat{j}_{\rm L} \hat{j}_{\rm R} \hat{\cal J}_{\rm 3L} 
\hat{\cal J}_{\rm 3R} \hat{\cal J}_{\rm 4L} 
\hat{\cal J}_{\rm 4R}\hat{J}_{\rm 1L}\hat{J}_{\rm 1R}
(-1)^{S_{\rm 1L}+S_{\rm 1R}}
\left\{ \begin{array}{ccc} \textstyle{\frac{1}{2}} & s_{\rm R} 
               & S_{\rm 1R} \\
             \textstyle{\frac{1}{2}}  &  S_2  & S_{\rm 1L}
\end{array}\right\}
\nonumber \\
&&\times 
\left\{ \begin{array}{ccc} l_{\rm L}   & s_{\rm L}   & j_{\rm L}   \\
  {\cal L}_{\rm L}  & \textstyle{\frac{1}{2}}  & {\cal J}_{\rm 3L} \\
                           L_{\rm 1L}   & S_{\rm 1L}  & J_{\rm 1L}
\end{array}\right\}
\left\{ \begin{array}{ccc} l_{\rm R}   & s_{\rm R}   & j_{\rm R}   \\
  {\cal L}_{\rm R}  & \textstyle{\frac{1}{2}}  & {\cal J}_{\rm 3R} \\
                           L_{\rm 1R}   & S_{\rm 1R}  & J_{\rm 1R}
\end{array}\right\}
\left\{ \begin{array}{ccc} L_{\rm 1L}   & S_{\rm 1L}   & J_{\rm 1L}   \\
  l_{z\rm L}  & \textstyle{\frac{1}{2}}  & {\cal J}_{\rm 4L} \\
                           L_2   & S_2  & J
\end{array}\right\}
\left\{ \begin{array}{ccc} L_{\rm 1R}   & S_{\rm 1R}   & J_{\rm 1R}   \\
  l_{z\rm R}  & \textstyle{\frac{1}{2}}  & {\cal J}_{\rm 4R} \\
                           L_2   & S_2  & J
\end{array}\right\}
\nonumber \\
&&\times \hat{L}'^2 (-1)^{L'}
\left\{ \begin{array}{ccc} l_{\rm R} & L_2 & L' \\
             l_{z\rm R}  &  {\cal L}_{\rm R}  & L_{\rm 1R}
\end{array}\right\}
\left\{ \begin{array}{ccc} l_{\rm R} & L_2 & L' \\
             l_{z\rm L}  &  l'  & L_{\rm 1L}
\end{array}\right\}
\left[
\hat{s}_{\rm L} \hat{s}_{\rm R}
  \left\{ \begin{array}{ccc} \textstyle{\frac{1}{2}} & \textstyle{\frac{1}{2}} 
               & s_{\rm R} \\
             \textstyle{\frac{1}{2}}  &  S_{\rm 1L}  & s_{\rm L}
\end{array}\right\}
\right.
\nonumber \\
&&\times (-1)^{l_{z\rm R}+L_{\rm 1L}}
\left( (-1)^{l_{z\rm L}}
\langle n' l' n_{z\rm L} l_{z\rm L} L' 
| n_{z\rm R} l_{z\rm R} {\cal N}_{\rm R} {\cal L}_{\rm R} L' 
\rangle_{\rm 8} 
\langle n_{\rm L} l_{\rm L} {\cal N}_{\rm L} {\cal L}_{\rm L} L_{\rm 1L} 
| n' l'  n_{\rm R} l_{\rm R} L_{\rm 1L} 
\rangle_{\rm 3} \right.
\nonumber \\
&&\left. 
+(-1)^{s_{\rm R}-s_{\rm L}
+{\cal L}_{\rm R}
-l_{\rm L}-{\cal L}_{\rm L}}
\langle n_{z\rm L} l_{z\rm L} n' l' L' 
| {\cal N}_{\rm R} {\cal L}_{\rm R} n_{z\rm R} l_{z\rm R}  L' 
\rangle_{\rm 8} 
\langle {\cal N}_{\rm L} {\cal L}_{\rm L} n_{\rm L} l_{\rm L} L_{\rm 1L} 
| n_{\rm R} l_{\rm R} n' l' L_{\rm 1L} 
\rangle_{\rm 3}\right)
\nonumber \\
&&\left. + \delta_{l_{\rm L},l_{\rm R}}\delta_{s_{\rm L},s_{\rm R}}
\delta_{{\cal N}_{\rm L},n'}
\delta_{{\cal L}_{\rm L},l'} (-1)^{{\cal L}_{\rm R}+l_{z\rm R}}
\langle n_{z\rm L} l_{z\rm L}  {\cal N}_{\rm L} {\cal L}_{\rm L} L' 
| {\cal N}_{\rm R} {\cal L}_{\rm R} n_{z\rm R} l_{z\rm R}  L' 
\rangle_{\rm 8} \right] \; ,
\end{eqnarray}
where 
$N_{\rm X}=2n_{\rm X}+l_{\rm X}+2{\cal N}_{\rm X}
+{\cal L}_{\rm X}+2n_{z \rm X}+l_{z \rm X}=
N_{1 \rm X}+2n_{z \rm X}+l_{z \rm X}, 
{\rm X}\equiv {\rm L,R}$. We will give more details on the derivation,
symmetry properties,
as well as a generalization that includes the isospin quantum
numbers, elsewhere \cite{NB99}. 

As the structure of the equation (\ref{fadfour}) is the same as
the Faddeev equation (\ref{Fadeq}), the discussion
of the spurious state problem for the three-fermion system
can be extended to the four-fermion system as well.
In particular, eigenstates of the operator ${\cal T}_4$ defined 
by the relation (\ref{metric4}) can be subdivided into two subspaces.
A physical subspace corresponding to the eigenvalue 4 spanned by totally
antisymmetric states and a spurious subspace spanned by eigenvectors
corresponding to the eigenvalue 0. In the basis of the ${\cal T}_4$
eigenstates, the Hamiltonian $\tilde{H}$ appearing in Eq. (\ref{fadfour})
has the same structure as shown in Eq. (\ref{hamstruct}). Its spurious
eigenstates can be identified by the action of ${\cal T}_4$ on the right
eigenstates, like in Eq. (\ref{anihil}), e.g., 
${\cal T}_4 |\psi_{(123)4};{\rm Sp}\rangle_{\rm K}=0$.

It is possible to symmetrize the Hamiltonian $\tilde{H}$ on the
physical subspace and eliminate the spurious state problem
at the same time. The symmetrized Hamiltonian then takes the form  
\begin{equation}
\bar{H}= H_0 + \bar{{\cal T}_4}^{1/2}\textstyle{\frac{1}{2}}
(V_{12}+V_{13}+V_{23})\bar{{\cal T}_4}^{1/2} \; , 
\end{equation}
where $\bar{\cal T}_4$ operates only on the physical subspace.

In Table \ref{tab2} we show the dimensions (D) of the basis (\ref{fourbas})
corresponding to particular $N=N_1+2n_z+l_z$
together with the number of physical states (Ph) and spurious states
(Sp) for the four-electron system with $J^\pi=0^+$. 
The relative number of physical states decreases with $N$ and approaches
about a fourth of all basis states for larger $N$.
In the last two rows of Table \ref{tab2} we present 
the ground state and the first excited state energies obtained 
in the calculations with the basis restricted by $N\le N_{\rm max}$,
where $N_{\rm max}$ corresponds to the number in the first row.
As in the three-electron calculations
the physical eigenenergies, which were obtained with 
the HO energy $\hbar\Omega=0.5$ a.u.,
are shown without the trivial center-of-mass contribution.
We note that the studied system, when described by Eq. (\ref{fadfour}), 
has a spurious ground state
with the energy 3.25 a.u., corresponding to the unperturbed
HO Hamiltonian $H_0$ (\ref{H0four}).
We note that the preliminary result of the ground-state energy 
of the studied system
obtained by the SVM is 5.6002 \cite{Varga2} 
(after subtracting 0.75 a.u. for the center-of-mass energy). 
As for the three-electron calculations, we get a fast convergence
to two-decimal place precision in our HO basis
calculation. 
A further improvement of the precision is slow and requires
larger model spaces. As in the three-electron calculation, 
a substantial
acceleration of convergence can be achieved by employing the
effective interaction approach \cite{NB98}. 
By replacing the interaction $V$ 
by a two-body effective interaction, we obtained a ground-state
energy of 5.5991 a.u. and a first-excited-state energy of 
5.6958 a.u. for $N_{\rm max}=14$. An additional speeding up 
of the rate of convergence
can be achieved by using the three-body effective interaction
\cite{NB99}.

\section{Conclusions}
\label{sec4}

To summarize, we have investigated the appearance of spurious states 
in the Faddeev formalism for few-body systems. Depending on the studied
system, such states may appear even among the lowest eigenstates.
We have also discussed how the spurious states can be identified
and how they can be eliminated by symmetrization of the 
Faddeev equations. We also noted that any improper truncation
in the treatment of the particle exchange operators 
(e.g., ${\cal T},{\cal T}_4$) may lead, in general, to the mixing of 
physical and spurious states.

We used three- and four-electron systems bound in an 
HO potential as an example. Due to the
repulsive character of the Coulomb interaction, such systems
have spurious ground states, when solved in the Faddeev
formalism. By examining the basis dimensions, we have illustrated 
that the number of physical states for the three-electron
system is about a third of the total number of basis states.
(It is exactly one-third, when the total number of basis states 
is divisible by three). For the four-electron system we formulated
equations for the Faddeev amplitudes antisymmetrized for the first 
three particles. Using this approach we observed that almost
$3/4$ of all the states were spurious.

\acknowledgements{
We would like to thank K. Varga for making available the SVM
results prior to publication.
One of the authors(W.G.) would like to thank Peter Schuck for drawing
his intention to possible occurrences of spuriosities in the Faddeev scheme. 
This work was partially supported by the NSF grant No. PHY96-05192.
P.N. also acknowledges partial support from the grant of the 
Grant Agency of the Czech Republic 202/96/1562.
B.R.B. would like to thank Takaharu Otsuka and the University of
Tokyo for their hospitality and to acknowledge the Japanese Society
for the Promotion of Science for fellowship support during
the latter part of this work.
}

\begin{table}
\begin{tabular}{cccccccccccc}
$N$ & 1 & 3 & 5 & 7 & 9 & 11 & 13 & 15 & 17 & 19 & 21  \\
\hline
 D  & 3 & 10& 21& 36& 55& 78 & 105& 136& 171& 210& 253\\ 
 Sp & 2 & 7 & 14& 24& 37& 52 & 70 & 91 & 114& 140& 169\\
 Ph & 1 & 3 & 7 & 12& 18& 26 & 35 & 45 & 57 & 70 & 84 \\
\hline
$E_1$& 3.4105 & 3.2787&3.2699&3.2669&3.2656&3.2649&3.2645&
3.2643&3.2641&3.2639&3.2638\\
$E_2$&   -    & 4.1002&4.0497&4.0472&4.0459&4.0453&4.0450&
4.0449&4.0448&4.0447&4.0447   
\end{tabular}
\caption{Results for three-electron system 
bound in a HO potential are presented. 
The first line shows the total of HO
quanta corresponding to the relative motion in the basis states.
The dimensions corresponding to the state $J^\pi=1/2^-$ are shown
in lines 2-4. The line 2 displays the full dimension (D), while
the lines 3 and 4 present the number of spurious (Sp) and physical 
(Ph) states, respectively. The lines 5 and 6 show calculated 
ground-state and first-excited-state energies, respectively. 
The calculations were done for $\hbar\Omega=0.5$
a.u. The ground-state center-of-mass energy (0.75 a.u.)
is not included in the energies shown. 
Energies corresponding to a particular $N$ were obtained in 
calculations, where all physical basis states up to $N$ were included. 
}
\label{tab1}
\end{table}

\begin{table}
\begin{tabular}{ccccccccc}
$N$ & 0 & 2 & 4 & 6 & 8 & 10 & 12 & 14 \\
\hline
 D  & 0 & 4 & 21& 68&178& 391& 767& 1390 \\
 Sp & 0 & 2 & 14& 48&128& 284& 564& 1024 \\
 Ph & 0 & 2 & 7 & 20& 50& 107& 203& 366  \\
\hline
$E_1$& - & 5.8829&5.6346&5.6094&5.6043&5.6022&5.6012&5.6006 \\
$E_2$& - & 6.0239&5.7396&5.7116&5.7034&5.7003&5.6988&5.6979 
\end{tabular}
\caption{Results for four-electron system 
bound in a HO potential are presented. 
The first line shows the total of HO
quanta corresponding to the relative motion in the basis states.
The dimensions corresponding to the state $J^\pi=0^+$ are shown
in lines 2-4. The line 2 displays the full dimension (D), while
the lines 3 and 4 present the number of spurious (Sp) and physical 
(Ph) states, respectively. The lines 5 and 6 shows calculated 
ground-state and first-excited-state energies, respectively.
The calculations were done for $\hbar\Omega=0.5$
atomic units. The ground-state center-of-mass energy (0.75 a.u.) 
is not included in the energies shown. 
Energies corresponding to a particular $N$ were obtained in 
calculations where all physical basis states up to $N$ were included.
}
\label{tab2}
\end{table}

\end{document}